\documentclass[aps,pre,superscriptaddress, twocolumn]{revtex4-1}
\usepackage[dvips]{graphicx}
\usepackage{amsmath}
\usepackage{amssymb}
\usepackage{natbib}
\usepackage{color}
\usepackage{epsfig}
\usepackage{subfigure}

\begin{document}

\title{Patterns of synchronization  in the hydrodynamic coupling of active colloids}

\author{Giovanni M. Cicuta}
\affiliation{Dipartimento  di Fisica, {Universit\`{a}} di Parma, Parco Area delle Scienze 7A, 43100 Parma, Italy }
\affiliation{INFN, Sez.di Milano-Bicocca, Gruppo Collegato di Parma}

\author{Enrico Onofri}
\affiliation{Dipartimento  di Fisica, {Universit\`{a}} di Parma, Parco Area delle Scienze 7A, 43100 Parma, Italy }
\affiliation{INFN, Sez.di Milano-Bicocca, Gruppo Collegato di Parma}

\author{Marco Cosentino Lagomarsino}
\affiliation{Genomic Physics Group, FRE 3214 CNRS ``Microorganism Genomics''}
\affiliation{University Pierre et Marie Curie, 15, rue de l'  Ecole de Medecine Paris, France}

\author{Pietro Cicuta}
 \affiliation{Cavendish Laboratory and
  Nanoscience Centre, University of Cambridge, Cambridge CB3 0HE,
  U.~K.}

 \email[correspondence to:]{giovanni.cicuta@fis.unipr.it}

\begin{abstract}
 A system of active colloidal particles driven by harmonic potentials to oscillate about the vertices of a regular polygon, with hydrodynamic coupling between all particles, is  described by a piece-wise linear model which exhibits various patterns of synchronization. Analytical solutions are obtained for this class of dynamical systems. Depending only on the number of particles, the synchronization occurs into states in which nearest neighbors oscillate either in-phase, or anti-phase, or in phase-locked (time-shifted) trajectories.
\end{abstract}

\maketitle

\section{Introduction}
 In passive colloidal particle systems, hydrodynamic interactions  determine the dissipative and dynamical behavior, rather than the real-space conformation; as such they influence the response properties rather than the equilibrium behavior.  However in actively driven systems that settle into a steady state, it can be the hydrodynamic interactions that determine the characteristics of the steady state.  For example, in  biological flows actuated  by cilia~\cite{gueron97},   the hydrodynamic coupling is possibly the most important feature involved in the synchronization of cilia beats~\cite{golestanian11b}.

In this paper we analyze theoretically a very simple dynamical system which was recently proven to adequately describe the motion of a system of colloidal spheres, maintained in oscillations by optical traps~\cite{cicuta10a,cicuta11z}. Motion occurs at low Reynolds number~\cite{brenner83}, and hydrodynamic forces between particles depend on the relative velocity~\cite{hunter81book,lauga09}.

The remarkable feature of the model is that the hydrodynamic interaction of the spheres leads to the synchronization of their phases, for any number of beads, for all initial conditions and for generic values of physical parameters (sphere size, viscosity of the fluid, stiffness of the forces). The synchronized state depends on the number of spheres, and on wether the optical traps provide attractive or repulsive forces. This produces a rich pattern of synchronized configurations.

The system of equations of motion is piece-wise linear. It has periodic solutions, which are linear combinations of the normal modes. Whilst we do not have a mathematical proof which would explain why this dynamical system always converges to a synchronized solution, we have performed a
 normal mode analysis and we show that based on this  it is possible to predict which synchronized solution is dominant. This analytical argument is in agreement with extensive numerical integration of the deterministic (no thermal noise) system,  and with the experiments and simulations  reported in the companion paper which include the effect of thermal noise~\cite{cicuta11z}.\\

Let us outline the physical system and its model, leading to the dynamic deterministic system of equation of motion which will be our main concern in the next section. We refer the reader to the literature which details a series of experiments and their analysis.
By focused laser beams, it was possible to confine effectively  an array of  micron-sized beads, each one into  a harmonic well. These colloidal particles interact exclusively through the hydrodynamic coupling. The collective fluctuation modes  resulting from hydrodynamic interactions have been studied in linear arrays \cite{quake06} and ring arrays \cite{ruocco07,cicuta10c}.   The ``geometric switch'' active driving model we study here was proposed in the context of cilia by~\cite{gueron97}; it was studied analytically for two beads and for infinite linear chains by Cosentino Lagomarsino and Bassetti in~\cite{bassetti03}, where they showed that an infinite linear chain of oscillators can sustain a traveling wave solution.  This model was then realized experimentally in \cite{cicuta10a}, by using optical tweezers to maintain a pair of  colloidal spheres   in oscillation by switching the
positions  of optical traps when a sphere reaches its limit position. This rule
leads to oscillations that are bounded in amplitude but free in
phase and period. The interaction between the oscillators is only
through the hydrodynamic flow induced by their motion; the colloidal particles in the experiment are subject to Brownian thermal fluctuations as would also be found in a biological context.  This system of two spheres in harmonic traps
 leads to  synchronized  motion in anti-phase \cite{cicuta10a}. A companion paper to this one
reports experimental results and the effect of noise, in the system of actively moving traps for many particles
\cite{cicuta11z}.  Very recently, Wollin and Stark~\cite{stark11}
     re-considered this model for the case of a chain of driven oscillators, performing numerical simulations  showing that for various potentials and forms of coupling the dynamic system evolves to a synchronized state which may be interpreted as a (discretized) wave propagating through the chain. The form of the wave depends on the driving forces and their being attractive or repulsive. This chain is closely related to the array of driven oscillators studied in this paper and experimentally observed~\cite{cicuta11z}. The main differences are the geometry (our oscillators are tangential to a ring) and the symmetry (in our model, the driving force acts in the same way in the two phases of each cycle, i.e. as each oscillator moves forward or backward).\\
     The periodic solutions analytically obtained in this paper may be interpreted as a wave propagating along the border of the ring, clockwise or anti-clockwise. Indeed if $x_j(t)$ describes the motion of the $j^{th}\mbox{-}$oscillator, our periodic solutions have the form $x_{j+1}(t+\Delta)=x_j(t)$, where $\Delta$ is a fixed,
     positive or negative, time interval, independent on $j$.
     In a recent paper \cite{golestanian11a} Uchida and Golestanian derived generic conditions on the driven oscillators to synchronize due to the hydrodynamic interactions. However, as it was already noted in \cite{stark11}, it is not clear if such conditions would apply to the ``geometric switch'' model.\\

 The paper is organized as follows: Section~\ref{model} outlines the model and the associated dynamical system. Section~\ref{analysis} shows the normal mode analysis that allows (in principle) a completely analytical determination of the periodic solutions of the system for any number of particles. Section~\ref{solutions} describes the periodic solutions, if the number of particles is $3,4,5$ and the dominant solution for any number of particles. In Section~\ref{repulsive} we replace the attractive forces with repulsive ones and obtain periodic solutions trivially related to the previous ones. \\

\begin{figure}[t!]
\centering
\includegraphics[bb = 320 375 633 850, height=50mm]{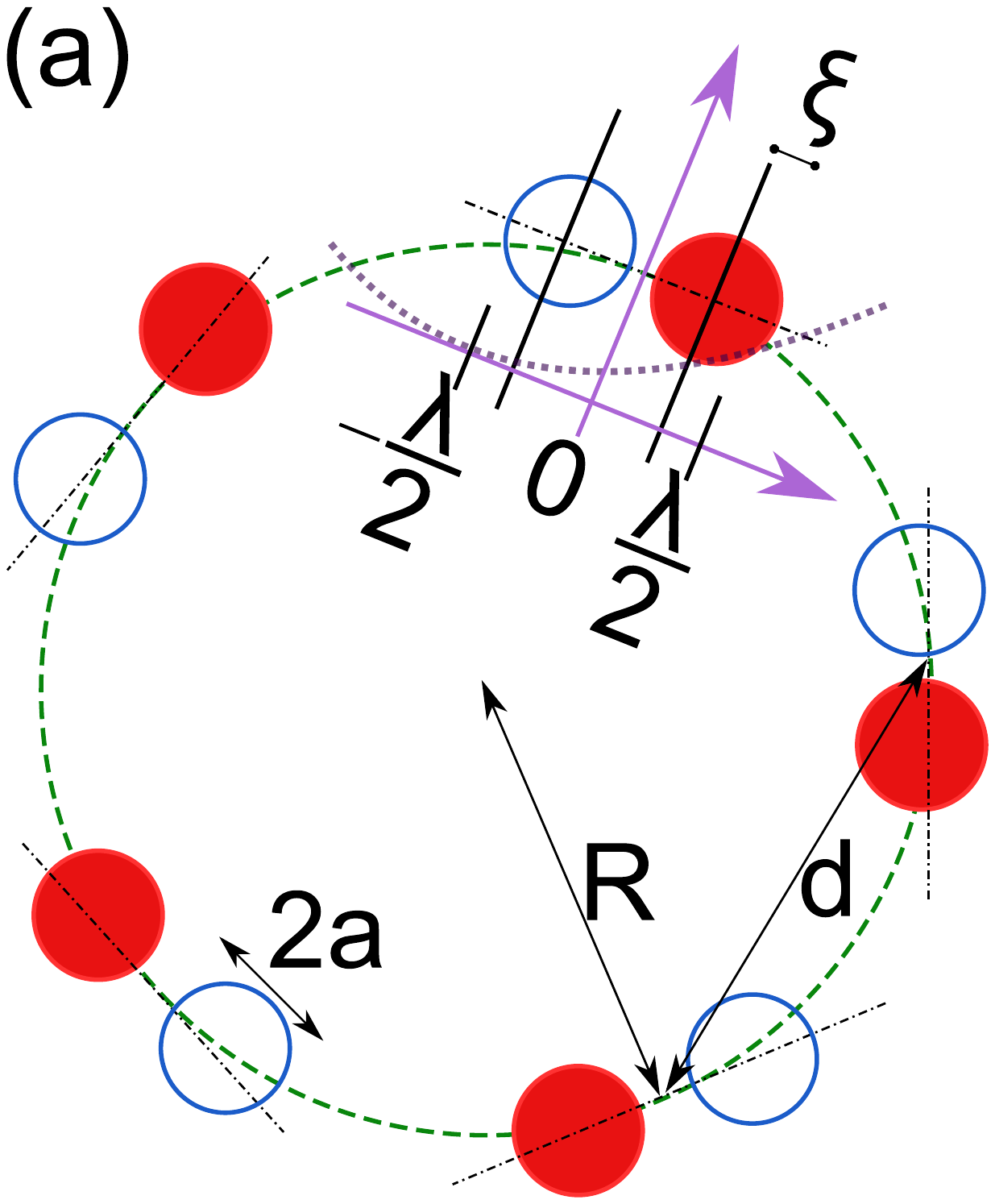}\\
\includegraphics[bb = 500 50 250 250, height=50mm]{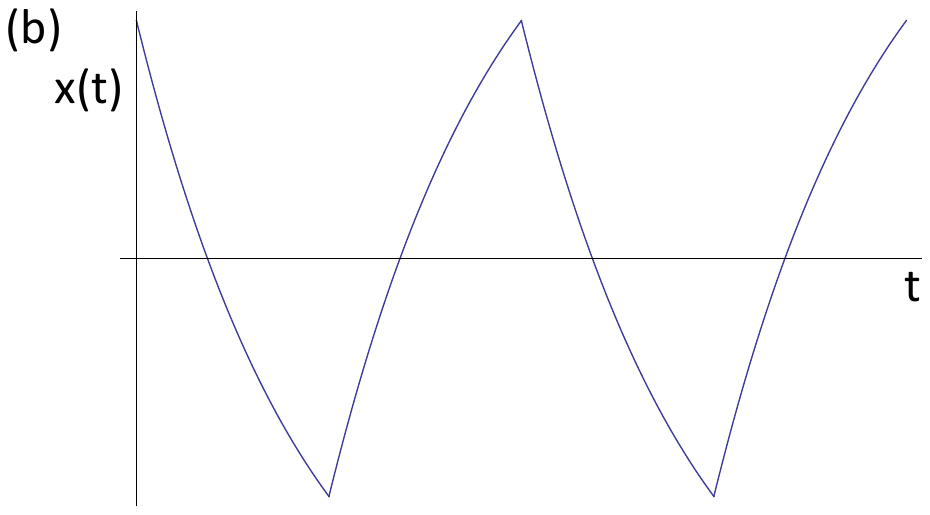}\\
 \caption{(Color online) (a) Diagram of the  ring arrangement with $n=6$ driven oscillators, illustrating the notation used in the main text. Colloidal particles  move tangential to the ring, with the center  of oscillations being the vertices of a regular $n-$polygon. The local tangential co-ordinate system is shown for one of the top particles. The particle configuration (solid disks) corresponds to an in-phase condition, when all 6 particles have just reached the geometric switch boundary. The particles are driven by harmonic traps towards the minimum of the potential, but they will reach first the other geometric switch boundary, when at the position indicated with open disks. This work also considers harmonic potentials with negative stiffness (i.e. repulsive) in section~\ref{repulsive}. (b) Trajectory of a driven oscillator with {\it attractive} driving force, uncoupled to other oscillators; $x(t)$ is measured along the tangential direction.}
  \label{fig1}
\end{figure}

\section{The model}\label{model}
Let us first describe the system in absence of hydrodynamic coupling between particles.
Fig.~\ref{fig1}(a) shows a ring of radius $R$, with $n$ spheres depicted at the vertices of a regular $n-$polygon. Each  vertex is the center of small oscillations of a particle bound to the ring.
The $j^{th}\mbox{-}$particle performs oscillations
about the position $R_j= R\,\theta_j^{(0)}=R\,\frac{2\pi}{n}(j-1)$, with $ j=1,2,\cdots , n$. On each side of this position, the amplitude of the oscillation is $\lambda/2-\xi$. This is assumed to be so small that we may replace the path on the arc with a segment of straight line, tangent to the ring, and then consider
   the motion of the $j^{th}\mbox{-}$particle on the tangent with a variable $x_j$:\,  $-(\lambda/2-\xi)\leq x_j(t)\leq \lambda/2-\xi$.
   The oscillations are driven by an attractive  harmonic potential (or repulsive harmonic potential, as discussed later in section~\ref{repulsive}) with center at $s=\pm \lambda/2$.
   In the hydrodynamic regime where Reynolds number is much smaller then unity, the inertial term may be neglected.

   An oscillation is set up that has fixed amplitude, but free phase and period. This is realized through a ``geometric switch'' condition. First the trap is set at $\lambda/2$, and particle  moves along the positive $x$ axis according to the differential equation
   $$\gamma_0{\dot x}(t)+\kappa\left( x(t)-\frac{\lambda}{2}\right)-f(x)=0,$$ where  $\kappa$ is the stiffness of the harmonic force, $\tau_0=\gamma_0/\kappa$ is the  relaxation time, and $f$ is a stochastic force due to thermal agitation of the fluid, which will be neglected in this work.
  As the particle reaches the position $\frac{\lambda}{2}-\xi$, the attractive harmonic potential is moved at the
   position $-\frac{\lambda}{2}$. The particle inverts its overdamped motion until it reaches the position
   $-\frac{\lambda}{2}+\xi$. At this time,  the potential jumps again.
   The resulting oscillatory motion (this is either a single bead, or a bead in a system with no coupling), neglecting any stochastic force, is depicted in Fig.~\ref{fig1}(b).
   The period $T_0$ of these uncoupled oscillations is $$\frac{T_0}{2\tau_0}\,=\,\log \frac{\lambda-\xi}{\xi}.$$\\

   In a system of $n$ beads, at any given time, the set of positions of the particles is a $n-$th dimensional vector ${\vec x}(t)=\{x_1(t), x_2(t),\cdots , x_n(t)\}$.
 In the absence of the hydrodynamic coupling, each particle would perform the same periodic  motion described above, a set of periodic overdamped
 relaxations. For any pair of particles, the ``phase differences'' $x_j(t)-x_k(t)$ would be set by initial conditions (typically some random values) and be constant in time.\\

   Let us now include the hydrodynamic interaction.
   The motion of the $j^{th}\mbox{-}$particle originates a force on the $i^{th}\mbox{-}$particle ${\vec f}_\alpha^{i,j}=(H^{-1})^{\alpha,\beta}_{i,j}v^{(j)}_\beta$, where $H$ is the Oseen tensor~\cite{brenner83}.
   In agreement with previous analysis~\cite{quake06,ruocco07,cicuta10c}
   we are led to the system of equations
 \begin{eqnarray}
  \left\{
\begin{array}{ccc}
&{\vec F}_i-\sum_{j=1}^n H^{-1}_{i,j} \frac{d {\vec r}_j(t)}{dt}+{\vec f}_i(t) =0  , \,\, i=1,2\cdots , n \qquad\\
   &{\vec r}_i(t)\cdot {\vec t}({\vec r}_i)=0   , \,\, {\vec t}(\theta)=\left(
   \begin{array}{cc}
   -\sin \theta \\ \cos \theta \end{array}\right).
   \end{array} \right.
\label{aa.1}
\end{eqnarray}

The deterministic force ${\vec F}_i$ acting on the $i$-th particle is a harmonic force, tangent to the ring, with the ``geometric switch'' rule, $${\vec F}_i=-\kappa\left[x_i(t)\mp \frac{\lambda}{2}\right]{\vec t}\left(\frac{2\pi (i-1)}{n}\right),$$ where $\pm \frac{\lambda}{2}$ is the coordinate of the bottom of the harmonic well.

 The   stochastic force $f_i(t)$ in Eq.~\ref{aa.1} represents the thermal noise on the $i$-th particle, and it can be assumed that
  $<f_i(t)>=0$, $<f_i(t_1)f_j(t_2)>=2 (\beta)^{-1} \, (H^{-1})_{ij}\, \delta (t_1-t_2)$, $\beta^{-1}=k_BT$ \cite{quake06}.
${\vec t}({\vec r}_i)$ is a versor tangent to the ring, at the position ${\vec r}_i$ , with anti-clockwise direction.
   For small oscillations, the Oseen tensor can be approximated by inserting the fixed distances among centers of oscillations, then:
\begin{eqnarray}
&&\gamma_0\,H^{\alpha \beta}_{ij}=\delta_{ij}\delta_{\alpha\beta}+(1-\delta_{ij})\frac{3a}{4r_{ij}}\left(\delta_{\alpha\beta}+\frac{r_{ij}^\alpha r_{ij}^\beta}{r^2_{ij}}\right),  \nonumber\\
&& \mathrm{where}\,\, \left( \begin{array}{cc}r_{jx}\\r_{jy} \end{array}\right) \simeq \left( \begin{array}{cc}R_{jx}\\R_{jy} \end{array}\right) =R
\left( \begin{array}{cc} \cos(j-1)2\pi/n\\ \sin(j-1)2\pi/n\end{array}\right)\,\nonumber\\
&&\mathrm{with} \,\, r_{ij}^\alpha \simeq R_j^\alpha-R_i^\alpha =-r_{ji}^\alpha, \,\mathrm{and} \,\, \alpha=1,2.
 \nonumber
 \end{eqnarray}

 We project the equations of the linearized Langevin-Oseen system
 along the tangents, replace the global coordinates of the $j^{th}\mbox{-}$particle ${\vec r}^{(j)}$ with the local one-dimensional coordinate $x^{(j)}$ and neglect the stochastic force, to obtain the deterministic linear system which holds in the time interval where no particle hits
   its left or right boundary $\pm( \frac{\lambda}{2}-\xi)$:
  \begin{eqnarray}
  \left(I+\frac{3a}{8R}C_n\right)\left( {\vec x}(t)-{\vec s}\right)+\tau_0 \frac{d}{dt} {\vec x}(t)=0,
 \label{cir.23x}
 \end{eqnarray}
 where ${\vec x}(t)$ is the $n$ component vector of the (local) position of the particles,
  ${\vec s} $ is the $n$-component vector of the (local) positions of the  minima of potential  proper for such time interval (i.e.  ${\vec s} $ 
  depends on the ${\vec x}(t)$ and on the previous history of the system),
  $C_n$ is a real symmetric circulant matrix of order $n$. The first row  of $C_n$ (which defines the entire matrix) is:
  \begin{eqnarray}
 C_n&=&\Bigg(
0 \, , \, \frac{\cos 2\pi/n+ \cos^2 \pi/n}{\sin \pi/n} \, , \,   \frac{\cos 4\pi/n+ \cos^2 2\pi/n}{\sin 2\pi/n} \, , \,  \cdots \, \nonumber \\
 \cdots &,& \,  \frac{\cos 2(n-1)\pi/n+ \cos^2 (n-1)\pi/n}{\sin (n-1)\pi/n}
  \Bigg).
\label{cir.12}
\end{eqnarray}

 \section{The deterministic dynamics}\label{analysis}

 We  now  study the deterministic system described by Eq.(\ref{cir.23x}).
 The hydrodynamic coupling  affects the overdamped oscillations, and determines the ``phase differences'' between particles. Most remarkably, it leads to  collective synchronized motions,  which as we will now show are best understood  by considering the   eigenstates of the Oseen coupling matrix $C_n$.

\begin{figure}[t!]
\centering
\includegraphics[width=80mm]{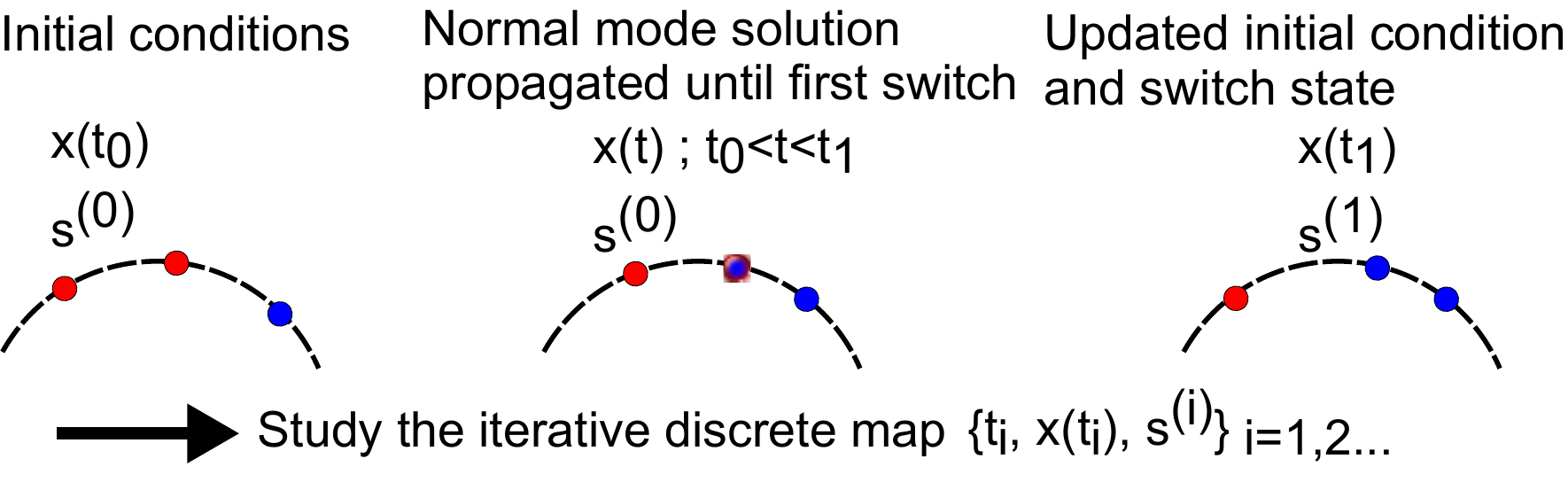}\\
 \caption{(Color online) A solution of the dynamical system can be obtained by iterating the linear evolution of the system, which is 
 simple in each of the eigenmodes, up to when any one of the beads reaches a  switch position. A periodic solution for a system of $N$ particles will return to the initial configuration (both particle and trap positions) after $2N$ switches.}
  \label{sketch}
\end{figure}

  Let us call $\lambda_j$, $j=1,\cdots , n$, the eigenvalues of the matrix $C_n$ arranged in increasing values, and
  ${\vec e}^{(j)} $ the corresponding normalized eigenvectors, $C_n\,{\vec e}^{(j)}=\lambda_j\,{\vec e}^{(j)}$. Let us call $h_j(t)=\left({\vec e}^{(j)}, {\vec x}(t)\right) $ the projection of the position vector on the eigenvectors, that is the normal modes of the linear system.
 Projecting the system of Eq.~\ref{cir.23x} onto each of the eigenvectors gives:
 \begin{eqnarray}
&&{\vec e}^{(j)}\cdot \left[  \left(I+\frac{3a}{8R}C_n\right)\left( {\vec x}(t)-{\vec s}\right)+\tau_0 \frac{d}{dt} {\vec x}(t)\right]=0,
 \label{cir.23xxa}
 \end{eqnarray}
  and one easily obtains the uncoupled  equations of motion for each normal mode:
  \begin{eqnarray}
	&&h_j(t)-\left(  {\vec e}^{(j)}\cdot {\vec s}\right)+\tau_j \frac{d}{dt}h_j(t)=0
 \nonumber\\
 &&\mathrm{with} \, \,\, \tau_j=\frac{\tau_0}{1+\frac{3a}{8R}\lambda_j}.
 \label{cir.23xx}
 \end{eqnarray}

 One sees that each normal mode $h_j(t)$ is related to a single exponential relaxation time $\tau_j$, and its linear differential equation is homogeneous, if and only if the term $\left(  {\vec e}^{(j)}\cdot {\vec s}\right)$ vanishes. In general, this is not the case.\\

  A  solution of the dynamical system can be obtained with the following simple strategy (Fig~\ref{sketch}): Given initial positions ${\vec x}(t_0)$ and initial configuration of potential ${\vec s^{(0)}}$, the evolution is trivially evaluated up to the time $t_1$ of the first hit. At this time one entry of the vector ${\vec s}$ changes sign, and
 the uncoupled system of Eq.~\ref{cir.23xx} is again evaluated with the new constant vector ${\vec s^{(1)}}$ up to the hit at time $t_2$. A basic role is then played by the continuity of each mode at the time of the hits: the final value before the hit becomes the new initial condition.\\

 The rotational discrete symmetry of the problem allows a detailed analytic study of the coupling matrix $C_n$. The most relevant features are the following:\\
 (1)~for every $n$ being even integer, the lowest eigenvalue $\lambda_1$ is a singlet, and its corresponding eigenvector is ${\vec e}^{(1)}=\frac{1}{\sqrt{n}}\left(1,-1,1,\cdots , -1\right) $; \\
(2)~for every $n$ being odd integer $n \geq 5$, the lowest eigenvalue $\lambda_1$ is a doublet. The two eigenvectors may be written as
  \begin{eqnarray}
 &&{\vec e}^{(1)}=\frac{1}{\sqrt{n}}\Big( 1, \cos (n-1)\pi/n, \cos 2(n-1)\pi/n,\cdots\nonumber\\
 &&\,\,\, \cdots \cos (n-1)^2\pi/n\Big),\nonumber\\
 &&{\vec e}^{(2)}=\frac{1}{\sqrt{n}}\Big( 0, \sin (n-1)\pi/n, \sin 2(n-1)\pi/n,\cdots \nonumber\\
 &&\,\,\, \cdots \sin (n-1)^2\pi/n\Big);
 \nonumber
  \end{eqnarray}
 (3)~for every $n$, the matrix $C_n$ has the eigenvector $ {\vec e}=\frac{1}{\sqrt{n}}\left(1,1,1,\cdots , 1\right) $. Only for $n=3$ this constant eigenvector corresponds to the lowest eigenvalue $\lambda_1$;\\
(4)~Since tr$[C_n]=0$, the lowest eigenvalue $\lambda_1<0 $, and therefore the relaxation time $\tau_1>\tau_0$ for every $n$. \\

\section{The analytic asymptotic solutions}\label{solutions}
The dynamics of the system depends on three combinations of parameters: $\tau_0=\gamma_0/\kappa$ sets the unit of time for the evolution between hits; $3a/(8R)$ measures the strength of the hydrodynamic interaction, and is limited by the bound $d=2R\sin \pi/n \gg 2a$. Finally $\xi/\lambda$, with $0<\xi/\lambda<1/2$, fixes the amplitude of the oscillations. The periodic solutions are here obtained for arbitrary values of the parameters (within the physical bounds given above). The plots for three and four particles, in Figures~\ref{fig2}, \ref{fig5}, \ref{fig12}, have $\xi/\lambda=1/4$ and  $3a/(8R)=\sqrt{3}/5$. The plots for five particles, in Figures~\ref{fig7} and \ref{fig9} have $\xi/\lambda=1/4$ and  $3a/(8R)=0.2$. These values of $3a/(8R)$ are much greater than sensible experimental  values~\cite{cicuta10a,cicuta11z}, but  are chosen here to spread the relaxation times $\tau_j$, so emphasizing the differences between periodic solutions.

\subsection{Case of  $n=3$}
 For $n=3$  the lowest eigenvalue of the coupling matrix $C_3$ is $\lambda_1=-\frac{1}{\sqrt{3}}$, a singlet. The next eigenvalue is a doublet $\lambda_2=\lambda_3=\frac{1}{2\sqrt{3}}$. The corresponding eigenvectors may be written:
 \begin{eqnarray}
  {\vec e}^{(1)}=\frac{1}{\sqrt{3}}\left(\begin{array}{ccc}1\\1\\1\end{array}\right)
  ,\,
  {\vec e}^{(2)}=\frac{1}{\sqrt{2}}\left(\begin{array}{ccc}0\\1\\-1\end{array}\right)
   ,\,
  {\vec e}^{(3)}=\frac{1}{\sqrt{6}}\left(\begin{array}{ccc} -2\\1\\1\end{array}\right).\nonumber\\
  \label{e.3}
 \end{eqnarray}

 We find the following periodic solutions for the system, here listed in order of decreasing duration of the periods:\\
 (a)~the three particles are synchronized in-phase;\\
 (b)~a pair of equivalent solutions, where the trajectories are phase-locked (more precisely, time-shifted).\\

  Let us consider first case (a), with the \textit{configuration of particles in-phase}, i.e. such that $x_1(t)= x_2(t)= x_3(t)$.  The vector of the positions of the potential is
 $${\vec s}=\pm \frac{\lambda}{2}\left(\begin{array}{ccc} 1\\1\\1 \end{array}\right).$$ The lowest normal mode is
  $h_1(t)=\frac{1}{\sqrt{3}}\left( x_1(t)+ x_2(t)+ x_3(t)\right)\to \sqrt{3}\, x_1(t)$ and it
  solves the equation:
   \begin{eqnarray}
&&   h_1(t)-\left(  {\vec e}^{(1)}\cdot {\vec s}\right)+\tau_1 \frac{d}{dt}h_1(t)=0 \, ,\nonumber\\
 &&\,\,\mathrm{where} \,\,\left(  {\vec e}^{(1)}\cdot {\vec s}\right)=\pm \sqrt{3}\frac{\lambda}{2}\, ,
 \mathrm{and} \,\, \tau_1=\tau_0\left(1- \frac{\sqrt{3}a}{8 R}\right)^{-1}. \nonumber
  \end{eqnarray}

Since $\left({\vec e}^{(j)}\cdot {\vec s}\right)=0$ for $j\neq 1$, the modes $h_2(t)$, $h_3(t)$ solve a homogeneous equation and  the choice $h_2(t)=h_3(t)=0$ is consistent.
As the three particles hit simultaneously a geometric switch border, all of  ${\vec s}\to -{\vec s}$. Therefore by continuity one obtains the opposite motion in the following time interval.
The three particles perform in-phase oscillations, satisfying the same equation as if they were un-coupled, as depicted on Fig.~\ref{fig1}(b),  but with the longer period $$  \frac{T_1}{2\tau_1}=\log \frac{\lambda-\xi}{\xi}.$$\\

 Let us now \textit{obtain the phase-locked solutions}, case (b) above. The symmetry of the problem, and results of numerical simulations, suggest to search for solutions where the time interval between hits is fixed, say $\Delta$. Assuming an initial time $t_0$ at one hit, the continuity of the trajectories requires:
 \begin{eqnarray}
 &&h_j(t_0+r\Delta)=\nonumber\\
 &&\,\,\,({\vec e}^{(j)}\cdot {\vec s}^{(r)})+\left[h_j(t_0+(r-1)\Delta)-({\vec e}^{(j)}\cdot {\vec s}^{(r)})\right]e^{-\Delta/\tau_j}, \nonumber\\
 &&\mathrm{with} \,\,\, r=1,2,3.
 \end{eqnarray}
 By assuming that the system has a period $T=6\Delta$, one obtains:
 \begin{eqnarray}
 &&h_j(t_0)\left(1+e^{-3\Delta/\tau_j}\right)=\nonumber\\
 &&\,\,\,\,=-  \left(1-e^{-\Delta/\tau_j}\right)\sum_{r=1}^3 \left({\vec e}^{(j)}\cdot {\vec s}^{(r)}\right)e^{-(3-r)\Delta/\tau_j}.
\nonumber
 \end{eqnarray}

\begin{figure}[t!]
\centering
\includegraphics[bb = 100 10 150 150, height=40mm]{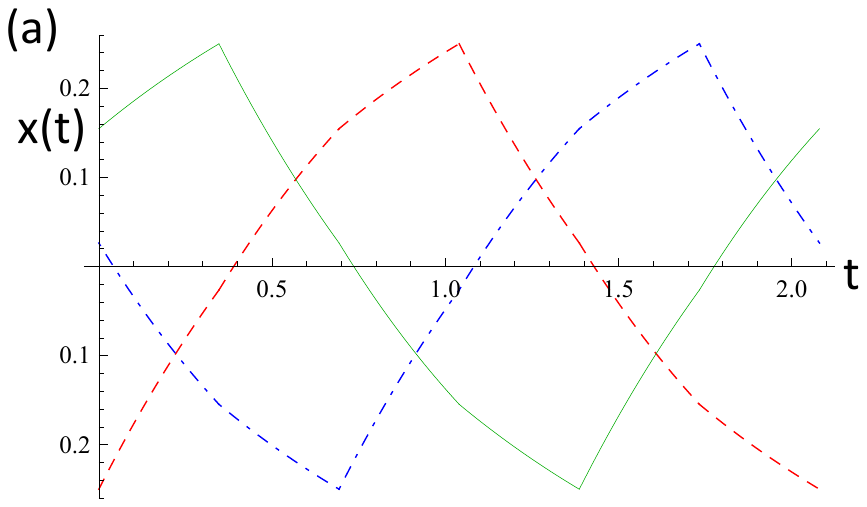}\\
\includegraphics[bb = 110 10 150 150, height=40mm]{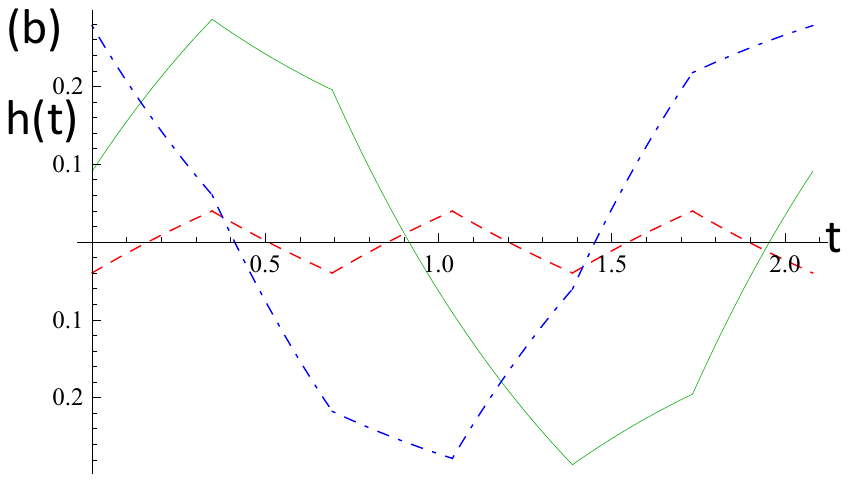}\\
 \caption{(Color online) The fastest (shortest period) solution for $n=3$, is phase-locked. (a) The trajectories of the particles 1 (dash-red), 2 (solid-green), 3 (dash/dot-blue) are plotted versus time, for one period.    (b) The normal modes $h_1(t)$ in red, $h_2(t)$ in green,  $h_3(t)$ in blue are plotted versus time.  Time is expressed in units of $\tau_0$. This solution is not stable, as can be shown by numerical simulation, see section~\ref{simulations}. The other solution for $n=3$ is the motion of all particles relaxing and switching in phase: it has longer period than the one illustrated here.}
  \label{fig2}
\end{figure}

 We choose $t_0$ when $x_1(t_0)=-\lambda/2+\xi$, the particle 2 is also moving toward $\lambda/2$ and particle 3 is moving in the opposite direction. That is we consider the sequence
 $\{ {\vec s}^{(j)} \}$, with  $j=1, \cdots , 6$ , $ {\vec s}^{(j+3)} = -{\vec s}^{(j)}$ , ${\vec s}^{(1)}={\vec s}^{(7)}$:
 \begin{eqnarray}
 \frac{2}{\lambda}{\vec s}^{(1)}= \left(\begin{array}{ccc}
 1\\1\\-1 \end{array}\right)\to \left(\begin{array}{ccc}
 1\\-1\\-1 \end{array}\right)\to \left(\begin{array}{ccc}
 1\\-1\\1 \end{array}\right)\to \nonumber\\  \to\left(\begin{array}{ccc}
 -1\\-1\\1 \end{array}\right)\to \left(\begin{array}{ccc}
 -1\\1\\1 \end{array}\right)\to \left(\begin{array}{ccc}
 -1\\1\\-1 \end{array}\right) =\frac{2}{\lambda}{\vec s}^{(6)}.
 \qquad \nonumber\\
 \label{e.1}
 \end{eqnarray}
 The initial conditions of this periodic solution are then
 \begin{eqnarray}
&&   h_1(t_0)\left(1+e^{-3\Delta/\tau_1}\right)=\nonumber\\
&&\,\,\,=-  \left(1-e^{-\Delta/\tau_1}\right)\frac{\lambda}{2\sqrt{3}} \left(  e^{-2\Delta/\tau_1}-e^{-\Delta/\tau_1}+1\right),
  \nonumber\\
 &&   h_2(t_0)\left(1+e^{-3\Delta/\tau_2}\right)=\nonumber\\
 &&\,\,\,=-  \left(1-e^{-\Delta/\tau_2}\right)\frac{\lambda}{\sqrt{2}} \left(  e^{-2\Delta/\tau_2}-1\right),
  \nonumber\\
 &&   h_3(t_0)\left(1+e^{-3\Delta/\tau_2}\right)=\nonumber\\
&&\,\,\, = \left(1-e^{-\Delta/\tau_2}\right)\frac{\lambda}{\sqrt{6}} \left(  e^{-2\Delta/\tau_2}+2e^{-\Delta/\tau_2}+1\right).
  \label{e.2}
 \end{eqnarray}

 Finally by inserting the initial condition $x_1(t_0)=\frac{1}{\sqrt{3}}h_1(t_0)-\frac{2}{\sqrt{6}}h_3(t_0)$ $=-\frac{\lambda}{2}+\xi$, we find the equation that fixes $\Delta$ in terms of the parameters of the problem. This is described in Appendix.
 Fig.\ref{fig2}a shows the trajectories of the particles 1, 2 and 3 as a function of time, for one period $T=6\Delta$. Time is expressed  in units of $\tau_0$.
 Fig.\ref{fig2}b plots the normal modes $h_1(t)$, $h_2(t)$ and $h_3(t)$, again for one period. The oscillations of the pair of normal modes corresponding to the degenerate eigenvalue have an amplitude much larger than the oscillations of the normal mode $h_1(t)$.  We verified that there exists a completely analogous phase locked synchronized solution where the role of particles $2$ and $3$ are exchanged. \\



\begin{figure}[t!]
\centering
\includegraphics[bb = 80 50 210 210, height=40mm]{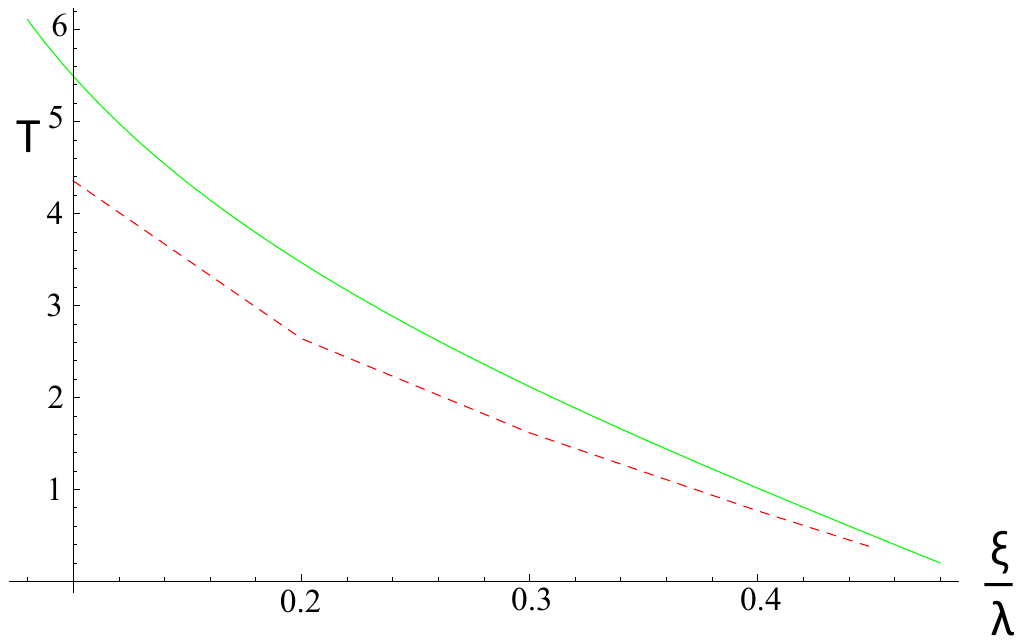}\\
 \caption{(Color online) The period of the phase-locking solution, dashed line, and the period of the solution with the $3$ oscillators in phase, in solid line, are plotted   versus $\xi/\lambda$. The periods, in units  of $\tau_0$, were evaluated with ${3a}/{(8R)}=\sqrt{3}/5$, that is $\tau_0/ \tau_1=0.8$.}
  \label{fig4}
\end{figure}

 In Fig.\ref{fig4}, we compare the period of the phase-locking solution, with the period of the solution with the $3$ oscillators in phase, as a function of $\xi/\lambda$.  One sees that the solution where the three particles are synchronized in-phase always has a longer period than the period of the pair of phase-locked  solutions, independently of the geometric conditions. This is also the solution which appears in the experiment and in the numerical simulations~\cite{cicuta11z}.

\subsection{Case of $n=4$}
 For $n=4$  the lowest eigenvalue of the matrix $C_4$ is a singlet  $\lambda_1=-1-\sqrt{2}$, the next is also a singlet  $\lambda_2=-1+\sqrt{2}$, and the next two are a doublet $\lambda_3=\lambda_4=1$. The eigenvectors may be written:
 \begin{eqnarray}
&&  {\vec e}^{(1)}=\frac{1}{2}\left(\begin{array}{ccc}1\\-1\\1\\-1\end{array}\right)
  , \,\,
  {\vec e}^{(2)}=\frac{1}{2}\left(\begin{array}{ccc}1\\1\\1\\1\end{array}\right),\nonumber\\
&&  {\vec e}^{(3)}=\frac{1}{\sqrt{2}}\left(\begin{array}{ccc} 0\\-1\\0\\1\end{array}\right)
  , \,\,
  {\vec e}^{(4)}=\frac{1}{\sqrt{2}}\left(\begin{array}{ccc} -1\\0\\1\\0\end{array}\right).
  \label{e.3v}
 \end{eqnarray}

  The periodic solutions of the system in order of decreasing periods are :\\
 (a)~adjacent particles  in anti-phase configuration;\\
 (b)~all particles  synchronized in-phase;\\
 (c)~the pair $(1,2)$ are anti-phase with the pair $(3,4)$. Also an equivalent solution where the pair $(1,4)$ are anti-phase with the pair $(2,3)$;\\
 (d)~a pair of phase-locked solutions, with the same period.\\

\begin{figure}[t!]
\centering
\includegraphics[bb = 100 10 150 150, height=40mm]{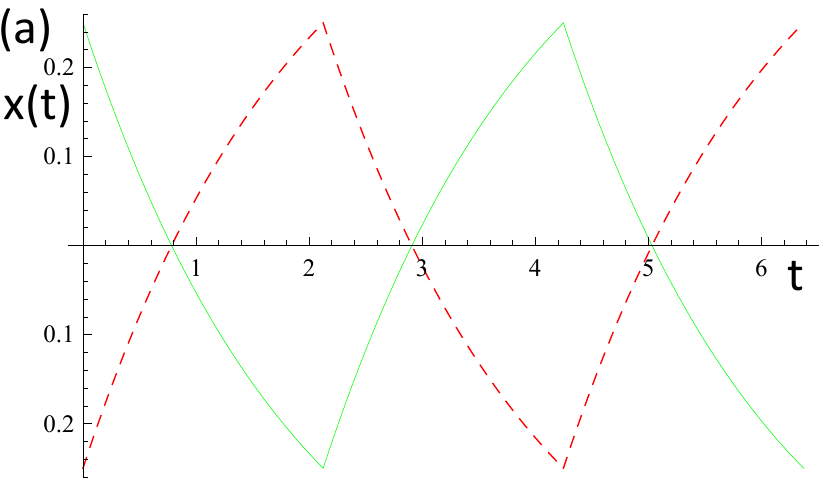}\\
\includegraphics[bb = 100 10 180 180, height=40mm]{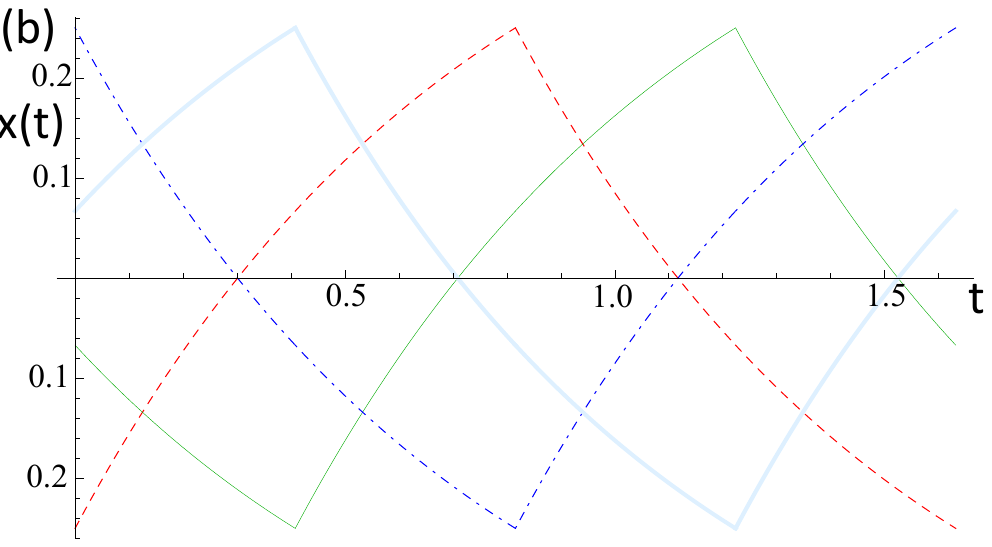}\\
 \caption{(Color online) (a) The ``slowest'' solution (longest period) for the case $n=4$ has nearest neighbors in antiphase with each other (hence next-nearest neighbors are in-phase).   The trajectories of the pair of particles $1$ and $3$ in red-dash, and the other pair in green-solid are plotted versus time, in units $\tau_0$. (b) There are two equally ``fastest'' solutions for  $n=4$. One has nearest-neighbor pairs in phase, and the other pairs in antiphase with each other. This would look like panel (a), but with particles 1,2 in red, and 3,4 in blue.  The other solution is shown in (b), and has  trajectories of the particles  $1$ to $4$, respectively in red-dash, green-solid, blue-dash/dot and cyan-thick are plotted versus time, in units $\tau_0$.}
  \label{fig5}
\end{figure}

 In the anti-phase configuration (a), the vector ${\vec s}$ of the positions of the potentials is proportional to the eigenvector ${\vec e}^{(1)}$. Then, all the normal modes $h_j(t)$ with $j=2,3,4$ obey homogeneous differential equations and may consistently vanish at all times. The normal mode  $h_1(t)$ generates the following solution:
 \begin{eqnarray}
 &&{\vec s}=\pm \frac{\lambda}{2}\left(\begin{array}{ccc} 1\\-1\\1\\-1 \end{array}\right) , \,\, \left({\vec e}^{(1)}\cdot {\vec s}\right)=\pm \lambda ,\nonumber\\
  &&  h_1(t)= \left({\vec e}^{(1)}\cdot {\vec x}(t)\right)=2\,x_1(t), \nonumber\\
 &&    {\vec x}_1(t)=-{\vec x}_2(t)={\vec x}_3(t)=-{\vec x}_4(t)  , \nonumber\\
   &&    h_1(t)-\left(  {\vec e}^{(1)}\cdot {\vec s}\right)+\tau_1 \frac{d}{dt}h_1(t)=0  , \nonumber\\
 && \mathrm{with}  \,\,\,\tau_1=\tau_0\left(1- \frac{(1+\sqrt{2})3a}{8R}\right)^{-1}. \nonumber
  \end{eqnarray}
  Here the four particles perform anti-phase oscillations, satisfying the same equation as un-coupled oscillators, but with $\tau_1$ replacing $\tau_0$. The period of this anti-phase solution is $$ T_1=2\tau_1\,\log \frac{\lambda-\xi}{\xi}.$$
 The trajectories for this case are shown in Fig.~\ref{fig5}a.

The next solution, case (b) above,  is the in-phase synchronization and may still be written in terms of just one normal mode, $h_2(t)$. Its relaxation time and period are
 \begin{eqnarray}&&\tau_2=\tau_0\left(1+ \frac{(\sqrt{2}-1)3a}{8R}\right)^{-1}  , \nonumber\\
  &&T_2={2\tau_2}\,\log \frac{\lambda-\xi}{\xi}<T_1. \nonumber
  \end{eqnarray}

 In the next solution, case (c),  the couple $x_1(t)=x_2(t)$ are in anti-phase with the couple $x_3(t)=x_4(t)$.
 The vector ${\vec s}$ of the position of the potential is proportional to the sum ${\vec e}^{(3)}+{\vec e}^{(4)}$.
 Then we may choose $h_1(t)=0$ , $h_2(t)=0$, and:\\
 \begin{eqnarray}
 &&{\vec s}=\pm \frac{\lambda}{2}\left(\begin{array}{ccc} 1\\1\\-1\\-1 \end{array}\right) , \,\,\left({\vec e}^{(3)}\cdot {\vec s}\right)=\left({\vec e}^{(4)}\cdot {\vec s}\right)=\mp \frac{\lambda}{\sqrt{2}} ,  \nonumber\\
 &&    {\vec x}_1(t)={\vec x}_2(t)=-{\vec x}_3(t)=-{\vec x}_4(t) =-\frac{1}{\sqrt{2}}h_4(t) , \nonumber\\
 &&  h_3(t)=h_4(t),   \nonumber\\
 && \tau_3=\tau_0\left(1+ \frac{3a}{8R}\right)^{-1} \, , \,\, T_3={2\tau_3}\,\log \frac{\lambda-\xi}{\xi}<T_2.
   \nonumber
  \end{eqnarray}
  The trajectories perform oscillations as depicted in Fig.~\ref{fig5}a, but the relaxation time  is replaced by $\tau_3$ and the pair of particles $1$ and $2$ in red, the other pair in blue.
 A completely equivalent solution has the pair of particles $x_1(t)=x_4(t)$  in anti-phase with the pair $x_2(t)=x_3(t)$.\\

 Finally there are phase-locked solutions, case (d) above, also with short periods $T$,  where $x_4(t)=x_1(t-3T/4)$, $x_3(t)=x_1(t-T/2)$, $x_2(t)=x_1(t-T/4)$. A completely equivalent periodic solution has the reverse order
 $x_4(t)=x_1(t+3T/4)$, $x_3(t)=x_1(t+T/2$, $x_2(t)=x_1(t+T/4)$.\\
  Since the sequence of the positions of the potentials is ${\vec s}=\frac{\lambda}{2}\left(1,-1,-1,1\right)\to -{\vec s} \to
  {\vec s} \to -{\vec s} \to \dots $, we find $h_1(t)=0$ , $h_2(t)=0$, and
  \begin{eqnarray}
&&  h_3(t_0)=\frac{\lambda}{\sqrt{2} }\frac{\left(1-e^{-\Delta/\tau_3}\right)^2}{1+e^{-2\Delta/\tau_3}}, \nonumber\\
&&  h_4(t_0)=\frac{\lambda}{\sqrt{2} }\frac{1-e^{-2\Delta/\tau_3}}{1+e^{-2\Delta/\tau_3}}, \nonumber\\
&&  \mathrm{with}\,\, \tau_3=\frac{\tau_o}{1+\frac{3a}{8R}}.
  \end{eqnarray}
  The period is:
  \begin{eqnarray}
 && T=4\, \Delta=2\tau_3\,\log \frac{\lambda-\xi}{\xi}=T_3,\nonumber\\
 &&\,\,\, \mathrm{where} \,\,  \xi=\lambda \frac{e^{-2\Delta/\tau_3}}{1+e^{-2\Delta/\tau_3}},
 \end{eqnarray}
same period as case (c),  and the motion is given by:
  \begin{eqnarray}
  &&x_1(t)=-\frac{1}{\sqrt{2}}h_4(t)   , \quad  x_2(t)=-\frac{1}{\sqrt{2}}h_3(t), \nonumber\\
  &&x_3(t)=\frac{1}{\sqrt{2}}h_4(t)   , \quad x_4(t)=\frac{1}{\sqrt{2}}h_3(t).     \nonumber
  \end{eqnarray}
 The trajectories for this are shown in Fig.~\ref{fig5}b,  for the particles $1$ to $4$.

\subsection{Case of $n=5$}
The odd-numbered systems, except $n=3$ discussed above, have general properties.
 For $n=5$  the lowest eigenvalue of the matrix $C_5$ is a doublet  $\lambda_1=\lambda_2$, the next is a singlet  $\lambda_3$, followed by a doublet $\lambda_4=\lambda_5$. The eigenvectors may be written
 \begin{eqnarray}
 &&\lambda_1=\lambda_2=-\frac{1}{2}\sqrt{29+\frac{22}{\sqrt{5}}}\simeq -3.116, \nonumber\\
 && \,\,\,{\vec e}^{(1)}=\sqrt{\frac{2}{5}}\left(\begin{array}{ccc}1\\ \cos \frac{4\pi}{5}\\ \cos \frac{2\pi}{5}\\ \cos \frac{2\pi}{5}\\   \cos \frac{4\pi}{5}\end{array}\right)
  , \,\,
  {\vec e}^{(2)}= \sqrt{\frac{2}{5}}\left(\begin{array}{ccc}0 \\  \sin \frac{4\pi}{5}\\ -\sin \frac{2\pi}{5}\\ \sin \frac{2\pi}{5}\\   -\sin \frac{4\pi}{5}     \end{array}\right), \nonumber\\
  && \lambda_3=\sqrt{13-\frac{22}{\sqrt{5}}}\simeq 1.778, \nonumber\\
  &&\,\,\,  {\vec e}^{(3)}=\sqrt{\frac{1}{5}}\left(\begin{array}{ccc} 1\\1\\1\\1\\1\end{array}\right), \nonumber\\
  &&\lambda_4=\lambda_5 \simeq 2.227  , \nonumber\\
   && \,\,\, {\vec e}^{(4)}= \sqrt{\frac{2}{5}}\left(\begin{array}{ccc} 1\\ \cos \frac{2\pi}{5}\\ \cos \frac{4\pi}{5}\\ \cos \frac{4\pi}{5}\\   \cos \frac{2\pi}{5}
  \end{array}\right)  , \,\,
  {\vec e}^{(5)}=\sqrt{\frac{2}{5}}\left(\begin{array}{ccc}0 \\ \sin \frac{2\pi}{5}\\ \sin \frac{4\pi}{5}\\ -\sin \frac{4\pi}{5}\\   -\sin \frac{2\pi}{5}      \end{array}\right).
  \nonumber
 \end{eqnarray}

  The periodic solutions of the system in order of decreasing periods are:\\
 (a)~a pair of phase-locked solutions, one with $x_1(t)=x_3(t+2T/10)=x_5(t+4T/10)=x_2(t+6T/10)=x_4(t+8T/10)$ and the analogous solution with opposite time-shifts;\\
 (b)~a pair of phase-locked solutions, one with $x_1(t)=x_2(t+2T/10)=x_3(t+4T/10)=x_4(t+6T/10)=x_5(t+8T/10)$ and the analogous solution with opposite time-shifts;\\
 (c)~a synchronized in-phase solution $x_j(t)=x(t)$ , $j=1,2,\cdots , 5$.\\
 The periods of these solutions are shown in Fig.~\ref{fig14}, and evaluated in the Appendix.\\

 \begin{figure}[t!]
\centering
\includegraphics[bb = 180 10 180 200, height=40mm]{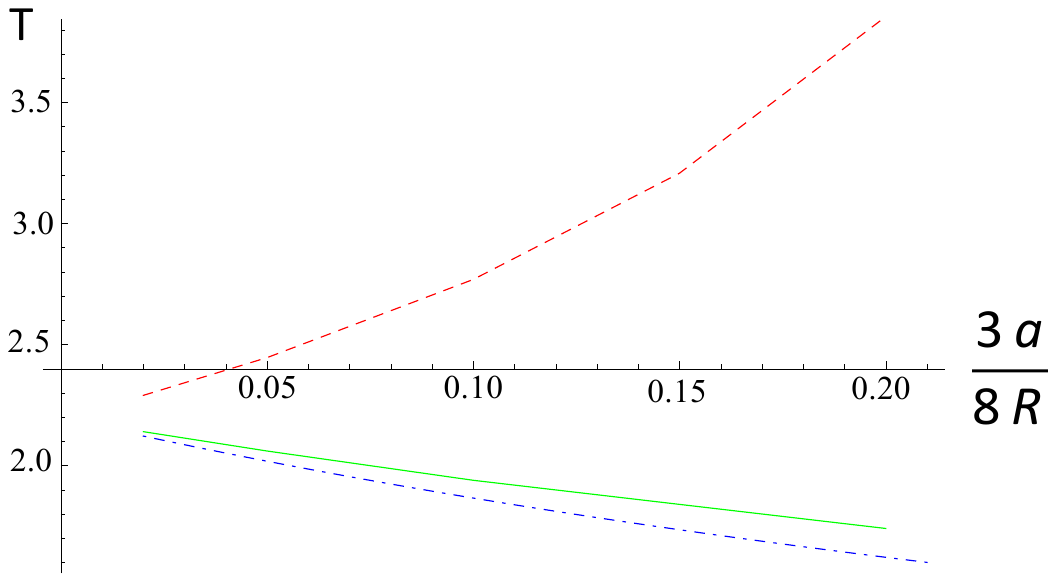}
 \caption{(Color online)  The periods    for the phase-shifted solution (red-dash) with adjacent trajectories $\{1,3,5,2,4 \}$,
 for the phase-shifted solution (green-solid) with adjacent trajectories $\{1,2,3,4,5 \}$, and for the in-phase solution (blue-dash/dot),in units of $\tau_0$, are plotted  versus the hydrodynamic coupling $3a/(8R)$. These periods are evaluated in the Appendix.}
  \label{fig14}
\end{figure}

 The phase-locked solution (a) is quite relevant since for every odd number of particles greater than $3$,  the system synchronizes into a periodic solution of this type.
    At any time, three particles are moving in one direction and $2$ are moving in the opposite direction. We call $\Delta$ the uniform time interval between any two consecutive hits. The period $T=10\, \Delta$.
 Take, for example, that at time $t_0$ the particles $1,3,5$ are moving towards the  positive direction, with $x_1(t_0)<x_3(t_0)<x_5(t_0)$. Then the sequence of the vectors
$\{ {\vec s}^{(j)} \}$, with $j=1, \cdots , 10$  , ${\vec s}^{(j)}=-{\vec s}^{(j+5)}$ is
 \begin{eqnarray}
&& \frac{2}{\lambda}{\vec s}^{(1)}= \left(\begin{array}{ccc}
 1\\-1\\1\\-1\\1 \end{array}\right)\to \left(\begin{array}{ccc}
 1\\-1\\1\\-1 \\-1\end{array}\right)\to \left(\begin{array}{ccc}
 1\\-1\\1 \\1\\-1\end{array}\right)\to \nonumber\\
 && \to \left(\begin{array}{ccc}
 1\\-1\\-1\\1\\-1 \end{array}\right)\to \left(\begin{array}{ccc}
 1\\1\\-1\\1\\-1 \end{array}\right)\to \left(\begin{array}{ccc}
 -1\\1\\-1 \\1\\-1\end{array}\right) = \frac{2}{\lambda}{\vec s}^{(6)}.\nonumber\\
 \label{e.50}
 \end{eqnarray}

 No eigenvector  ${\vec e}^{(j)}$ is orthogonal to all of the eigenvectors $\{ {\vec s}^{(j)} \}$ of this sequence, therefore all the five normal modes $h_j(t)$ contribute to the solution.\\

 The continuity and periodicity conditions fix the initial conditions:
 \begin{eqnarray}
 &&h_j(t_0)\left(1+e^{-5\Delta/\tau_j}\right)=\nonumber\\
 &&\,\,=-\left(1-e^{\Delta/\tau_j}\right)\sum_{r=1}^5\left({\vec e}^{(j)}\cdot
 {\vec s}^{(r)}\right) \,e^{-(5-r)\Delta/\tau_j} = \nonumber\\
 &&\,\,=-\frac{\lambda}{2}\,{\vec e}^{(j)}\cdot \left(
 \begin{array}{cccccccc} 1-e^{-5\Delta/\tau_j}\\ 1-2e^{-\Delta/\tau_j}+e^{-5\Delta/\tau_j}\\
 -1+2e^{-2\Delta/\tau_j}-e^{-5\Delta/\tau_j}\\
 1-2e^{-3\Delta/\tau_j}+e^{-5\Delta/\tau_j}\\
 -1+2e^{-4\Delta/\tau_j}-e^{-5\Delta/\tau_j}\end{array}\right)\, , \nonumber\\
 &&{\rm and} \,\, {\rm as}\,\, {\rm usual} \,\,\tau_j=\frac{\tau_0}{1+\frac{3a}{8R}\lambda_j}.
 \label{e.51}
 \end{eqnarray}

  One last equation determines the time interval $\Delta$. If we use $x_1(t_0)=-\frac{\lambda}{2}+\xi$, it is
 \begin{eqnarray}
 -\frac{\lambda}{2}+\xi=\sqrt{ \frac{2}{5}}\left(h_1(t_0)+h_4(t_0)\right)+\sqrt{ \frac{1}{5}}h_3(t_0).
 \label{e.52}
 \end{eqnarray}

\begin{figure}[t!]
\centering
\includegraphics[bb = 130 5 180 200, height=40mm]{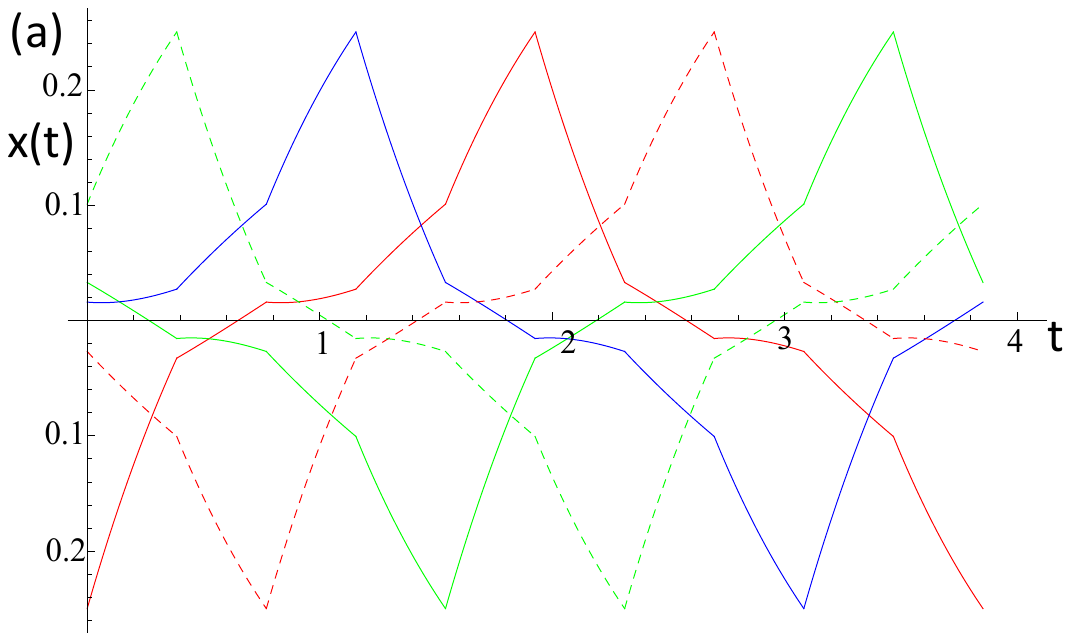}\\
\includegraphics[bb = 130 5 180 200, height=40mm]{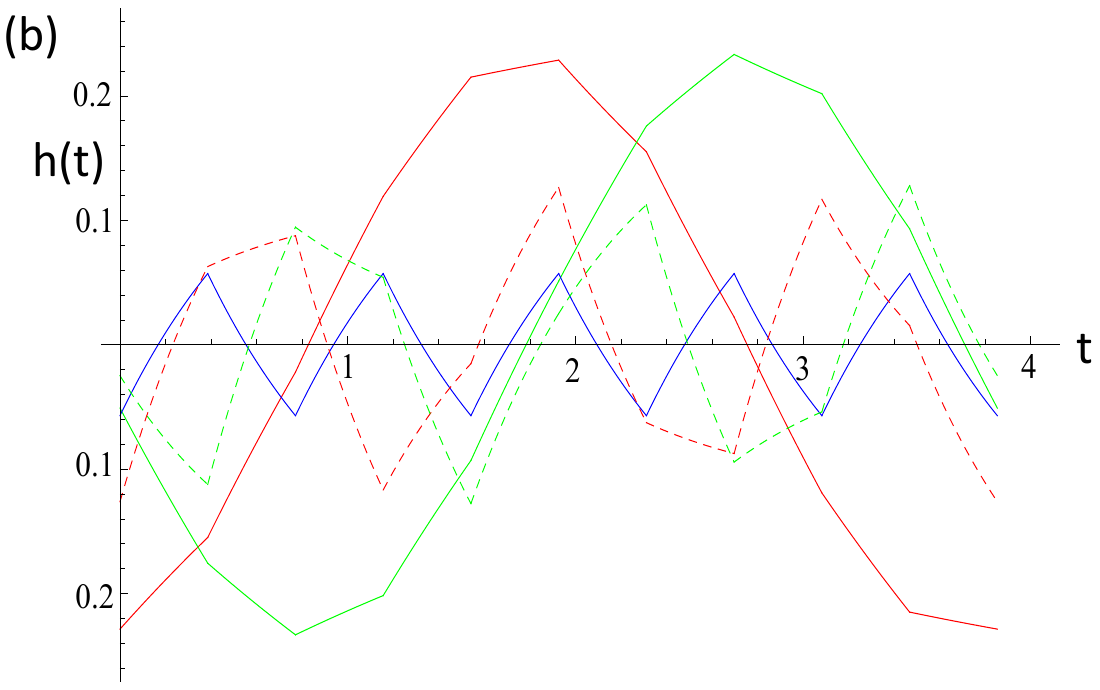}\\
 \caption{(Color online) This phase locked solution for $n=5$, with the nearest neighbors almost in anti-phase,  is the solution with the longest period, and is the stable state with attractive potentials. (a) The trajectories of the particles $1$ to $5$, respectively in red, green, blue, red-dash and green-dash are plotted versus time, in units $\tau_0$. To draw the figure, the coupling was chosen ${3a}/({8R})=0.2$, then the period of the solution is $T \simeq 3.8547 \tau_0$. (b) The normal modes $h_j(t)$ , $j=1,2,\cdots 5$, respectively in red, green, blue, red-dash and green-dash are plotted versus time, in units $\tau_0$. }
  \label{fig7}
\end{figure}

 The trajectories are shown in Fig.~\ref{fig7}a,  for the particles $1$ to $5$ (the coupling is chosen at ${3a}/({8R})=0.2$, and  then the period of the solution is $T \simeq 3.8547 \tau_0$).\\

  Fig.~\ref{fig7}b shows the complex behaviour of normal modes $h_j(t)$ , $j=1,2,\cdots 5$.
 The doublet of normal modes $1$ and $2$ present   oscillations with an amplitude much larger than the oscillations of the
 doublet of normal modes $4$ and $5$. The singlet normal mode $3$, related to the center of mass $\sum_{j=1}^5 x_j(t)$, has even smaller oscillations.

 Another pair of phase-locked solutions, case (b), with shorter period has the trajectories of particles $\{1,2,3,4,5\}$ and the equivalent solution with the arrangement of trajectories $\{1,5,4,3,2\}$. Here, like in case (a), at any time there are three particles moving in one direction and $2$ moving in the opposite direction. We call $\Delta$ the uniform time interval between two hits. The period $T=10\, \Delta$.
 Taking for example that at time $t_0$ the particles $1,4,5$ are moving toward the positive direction, with $x_1(t_0)<x_5(t_0)<x_4(t_0)$, then the sequence of the vectors
$\{ {\vec s}^{(j)} \}$, with $j=1, \cdots , 10$  , ${\vec s}^{(j)}=-{\vec s}^{(j+5)}$ is
 \begin{eqnarray}
&& \frac{2}{\lambda}{\vec s}^{(1)}= \left(\begin{array}{ccc}
 1\\-1\\-1\\1\\1 \end{array}\right)\to \left(\begin{array}{ccc}
 1\\-1\\-1\\-1 \\1\end{array}\right)\to \left(\begin{array}{ccc}
 1\\1\\-1 \\-1\\1\end{array}\right)\to \nonumber\\
 && \to \left(\begin{array}{ccc}
 1\\1\\-1\\-1\\-1 \end{array}\right)\to \left(\begin{array}{ccc}
 1\\1\\1\\-1\\-1 \end{array}\right)\to \left(\begin{array}{ccc}
 -1\\1\\1 \\-1\\-1\end{array}\right) = \frac{2}{\lambda}{\vec s}^{(6)}.\nonumber\\
 \label{e.53}
 \end{eqnarray}

As before, the continuity and periodicity conditions fix the initial conditions:
 \begin{eqnarray}
 && h_j(t_0)\left(1+e^{-5\Delta/\tau_j}\right)=\nonumber\\
 &&\,\,=-\left(1-e^{\Delta/\tau_j}\right)\sum_{r=1}^5\left({\vec e}^{(j)}\cdot
 {\vec s}^{(r)}\right) \,e^{-(5-r)\Delta/\tau_j} = \nonumber\\
 &&\,\,\,=-\frac{\lambda}{2}\,{\vec e}^{(j)}\cdot \left(
 \begin{array}{cccccccc}
  1-e^{-5\Delta/\tau_j}\\
    1-2e^{-3\Delta/\tau_j}+e^{-5\Delta/\tau_j}\\
 1-2e^{-\Delta/\tau_j}+e^{-5\Delta/\tau_j}\\
-1+2e^{-4\Delta/\tau_j}-e^{-5\Delta/\tau_j}\\
   -1+2e^{-2\Delta/\tau_j}-e^{-5\Delta/\tau_j}
 \end{array}\right), \nonumber\\
 &&{\rm and} \,\, {\rm as} \,\,{\rm usual}\,\,\tau_j=\frac{\tau_0}{1+\frac{3a}{8R}\lambda_j}.
 \label{e.55}
 \end{eqnarray}

\begin{figure}[t!]
\centering
\includegraphics[bb = 130 5 180 200, height=40mm]{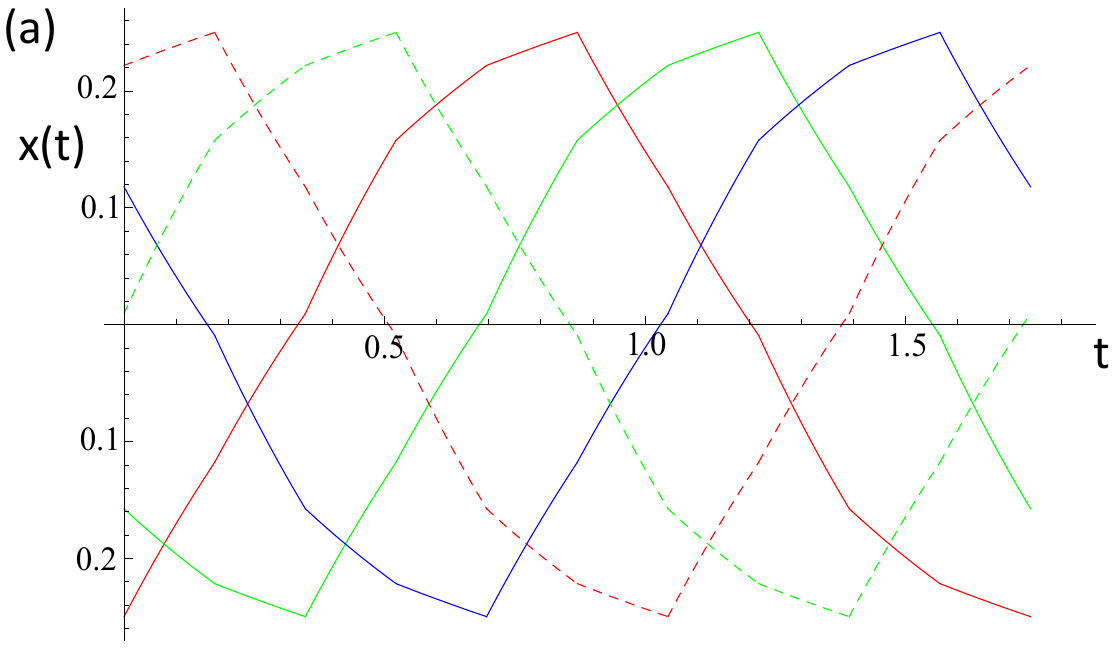}\\
\includegraphics[bb = 130 5 180 200, height=40mm]{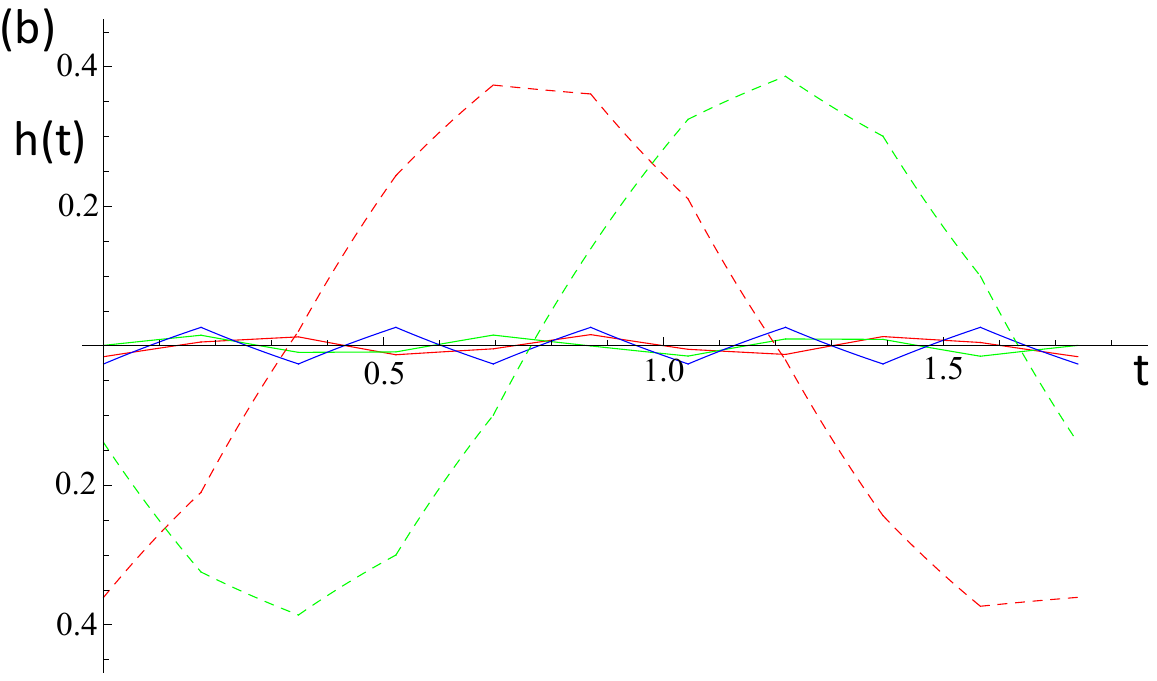}\\
 \caption{(Color online) This phase-locked solution has sorter period than the one depicted in Fig~\ref{fig7}. (a) The trajectories of the particles $1$ to $5$, respectively in red, green, blue, red-dash and green-dash
 are plotted versus time, in units $\tau_0$. To draw the figure, the coupling ${3a}/({8R})=0.2$, then the period of the solution is $T \sim 1.7407 \tau_0$. (b) The the normal modes $h_j(t)$ ,$j=1,2,\cdots 5$ , respectively in red, green, blue, red-dash and green-dash
 are plotted versus time, in units $\tau_0$. To draw the figure, the coupling ${3a}/({8R})=0.2$.}
  \label{fig9}
\end{figure}

 By using the same initial condition, $x_1(t_0)=-\frac{\lambda}{2}+\xi$, the equation that determines $\Delta$ still has the form of Eq.~(\ref{e.52}).
The trajectories are shown in Fig.~\ref{fig9}a,  for the particles $1$ to $5$. In the figure, the coupling is ${3a}/({8R})=0.2$, and  then the period of this solution is $T \sim 1.7407 \tau_0$.\\

Fig.~\ref{fig9}b shows the normal modes $h_j(t)$ ,$j=1,2,\cdots 5$.
 The doublet of normal modes $4$ and $5$, which have the shortest relaxation time, oscillate with much larger amplitudes than either  the
 doublet of normal modes $1$ and $2$ or  the singlet normal mode $3$.\\

The synchronized in-phase solution, case (c),  for the five particles is very simple, since   only the mode $h_3(t)$ contributes, whilst all the $h_j(t)=0$ for $j\neq 3$.  The
 relaxation time and period are
  \begin{eqnarray}
   \tau_3=\tau_0\left(1+ \frac{3a}{8R}\lambda_3\right)^{-1} , \,\,\,T_3={2\tau_3}\,\log \frac{\lambda-\xi}{\xi}.
   \end{eqnarray}
This is the ``fastest'' periodic solution.

\subsection{Case of higher $n$}
As the number of particles of the system increases, the analytic periodic solutions of the system become more and more complex. Yet some few simple features are generic.\\

We already mentioned that for every particle number $n$ being an \emph{even integer}, the lowest eigenvalue $\lambda_1$ is a singlet, its
eigenvector   is $ {\vec e}^{(1)}=\frac{1}{\sqrt{n}}\left(1,-1,1,\cdots , -1\right)$.
This allows a periodic solution where adjacent particles are in anti-phase, just as the first solution (a) described for the system of four particles. Since $h_1(t)$ is the unique normal mode non-vanishing,
 the period of the anti-phase solution is
 $$ T_1=2\tau_1\,\log \frac{\lambda-\xi}{\xi},\,\,
 \tau_1=\tau_0  \left(1+\frac{3a}{8R}\lambda_1\right)^{-1},\,\,\, {\rm where}$$
 $$\lambda_1=(-1)^{n/2+1}-2\sum_{j=1}^{n/2-1}(-1)^j\frac{ \cos \frac{2j\pi}{n}+\cos^2\frac{j\pi}{n}}{\sin \frac{j\pi}{n}}  \,\, {\rm that} \,\, {\rm is,}$$
 in particular:
    $$   \lambda_1\sim  -2.414\, \, (n=4);$$
  $$\lambda_1 \sim -4.577\, \, (n=6);$$
$$\lambda_1 \sim -6.528\,\, (n=8). $$
This anti-phase solution is the periodic solution with the longest period, for every  even $n$.\\

We also mentioned that for every $n$ being odd integer $n \geq 5$, the lowest eigenvalue $\lambda_1$ is a doublet.  The periodic solution with the longest period is analogous to the solution (a) for the system of five particles. That is   a pair of equivalent solutions where a sequence of trajectories are time shifted by a uniform time interval.
For $n=7$ the pair of sequences is $\{1,3,5,7,2,4,6\}$ and $\{1,6,4,2,7,5,3\}$. One may notice that if $n>>1$, adjacent particles are time shifted by almost half period, that is are approximately in anti-phase. Thus the distinction between even and odd number $n$ decreases, as $n$ increases.\\

\subsection{Simulations}\label{simulations}

To verify which of the analytical solutions is the dynamical steady
state, we resorted to numerical integration.  Two different codes have
been developed. 

Firstly we solved the coupled system of Eq.\ref{cir.23x} which 
corresponds to the tangential motion, using the solver {\tt ode113}
of {\tt matlab}, which implements the Adams-Bashforth-Moulton
multi-step  (i.e. timestep optimizing) method. We used the event handling facility integrated into all
{\tt matlab's} ode solvers to intercept the instant of time at which
any one of the particles hits the switch position. 
As a stabilizing
measure, we imposed that at a ``hit'' all particles which are in the
range of $\delta s$ from their switches are considered to have hit their
target and consequently their centers of force switch their
positions; several values in the interval $10^{-8}\,\lambda\le \delta s \le
10^{-5}\,\lambda$ have been used with no substantial difference on the results.

Secondly, a different code  described in \cite{ermak1978,cicuta10a,cicuta11z} with fixed timestep, is also used, enabling testing in the absence of noise but also the robustness at finite temperature. As well as exploring noise, this Brownian Dynamics  (BD) code  solves a model which is very close to the experimental condition~\cite{cicuta11z}: the colloidal particles are free to move in a two dimensional environment, and the tangential and normal forces can be set independently at different stiffnesses. Finally, experiments typically work at a finite sampling rate, and this can be mimicked closely in the fixed-timestep BD code as was discussed in previous study~\cite{cicuta10a}.

 The numerical solution (both the simulation approaches described above) show the remarkable result that for every particle number $n$ and every initial condition, the system evolves to a synchronized configuration corresponding to the periodic solution with the longest period  (for the case of positive trap stiffness $\kappa>0$ discussed up to here). If such a configuration is a pair of equivalent solutions, as it happens for odd $n \geq 5$, the synchronization occurs on either one of the pair.
These findings also agree with the experiment in~\cite{cicuta11z}.

Another result of the extensive numerical solutions is that we never observe different solutions from the cases that were investigated analytically (in principle, we could not have ruled out the presence of much more complex and less symmetric solutions).

  \section{Repulsive potential}\label{repulsive}

  It is interesting to consider  a harmonic {\it repulsive} potential, $\kappa<0$, acting  on the colloidal spheres. This may be realized experimentally by tailoring an appropriate potential landscape with time-sharing or holographical optical traps.   Then the synchronization of the system of oscillating particles due to the hydrodynamical interaction still occurs, but on a periodic configuration different from the one observed in the attractive case.
  With minimal changes from the previous analysis we can describe the periodic solutions occurring in the case of repulsive harmonic potentials. Again the numerical simulations of this case  confirm that the systems converge onto one of the analytical solutions, and in this case it is one with short period.\\

\begin{figure}[t!]
\centering
\includegraphics[bb = 130 5 180 200, height=40mm]{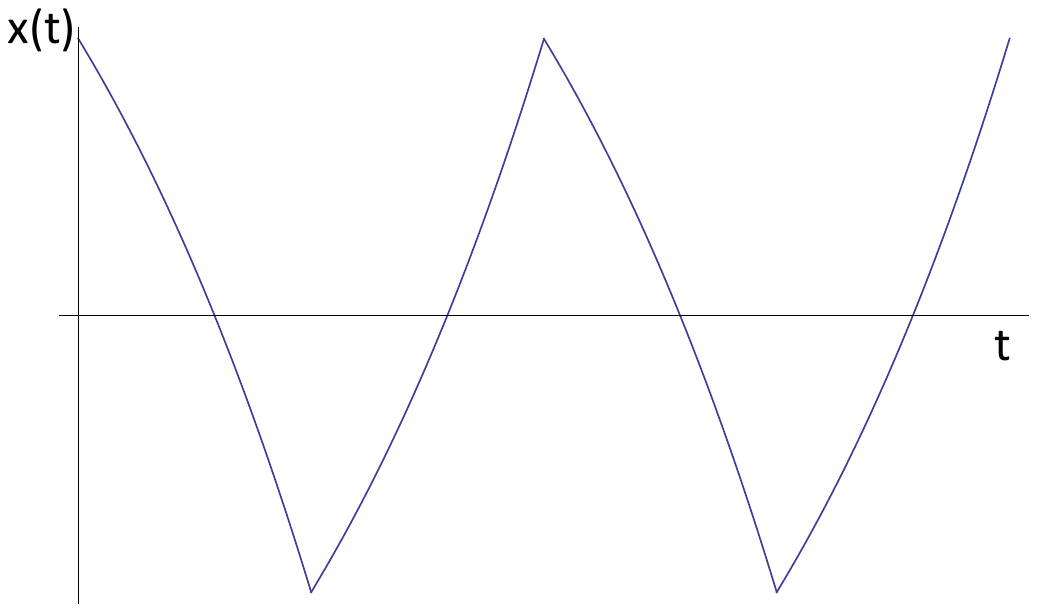}\\ \caption{(Color online) Trajectory of a driven oscillator with {\it repulsive} driving force, uncoupled to other oscillators. Note the increase in velocity between switch positions, characteristic of the motion with potentials of negative stiffness.}
  \label{fig11}
\end{figure}

With negative stiffness, if we ignore the hydrodynamic coupling, a particle moves toward the positive $x$
axis according to the differential equation $\gamma_0 \, {\dot x}(t)+\kappa\left( x(t)+\frac{\lambda}{2}\right)-f(x)=0$,
 where now $\kappa<0$ is the stiffness of the \emph{repulsive} harmonic force, $\tau_0 =\gamma_0/|\kappa|$,
 $f$ is a stochastic force due to thermal agitation of the
fluid, which will be ignored in the analysis of the deterministic system. In this case, the motion is exponentially \emph{accelerated} until the particle reaches the position $\frac{\lambda}{2}-\xi$. At that time
  the center of the repulsive harmonic potential
moves from the position $-\frac{\lambda}{2}$ to   $\frac{\lambda}{2}$. The particle inverts its  motion until
it reaches the position $-\frac{\lambda}{2}+\xi$ when the potential jumps again. The resulting
oscillatory motion for a single particle (or one in a non-coupled system), neglecting the stochastic force, is depicted in Fig.~\ref{fig11}.\\

 The period $T_0$ of these uncoupled oscillations is $T_0=2\tau_0\,\log \frac{\lambda-\xi}{\xi}$.
 The oscillatory motion is similar to the attractive potential case, shown in Fig.~\ref{fig1}, except for the concavity.\\

 One may repeat the analysis of the attractive case, and obtain the linear deterministic system, valid in a time interval between two hits:
  \begin{eqnarray}
  \left(I+\frac{3a}{8R}C_n\right)\left( {\vec x}(t)-{\vec s}\right)-\tau_0 \frac{d}{dt} {\vec x}(t)=0,\,\,
	\tau_0=\frac{\gamma_0}{|\kappa|},\nonumber\\
 \label{cir.23xy}
 \end{eqnarray}
 where ${\vec x}(t)$ is the $n$ component vector of the (local) position of the particles,
  ${\vec s} $ is the $n$ component vector of the (local) position of the  minima of the repulsive potential  proper for such time interval (opposite of the previous case),
  $C_n$ is the same real symmetric circulant matrix of order $n$, given in eq.(\ref{cir.12}).
  We again define the normal modes $h_j(t)=\left( {\vec e}^{(j)},{\vec x}(t)\right)$ and their decoupled equations:
  \begin{eqnarray}
h_j(t)-\left(  {\vec e}^{(j)}\cdot {\vec s}\right)-\tau_j \frac{d}{dt}h_j(t)=0,
\,\, \tau_j=\frac{\tau_0}{1+\frac{3a}{8R}\lambda_j}. \nonumber\\
 \label{cir.23xxy}
 \end{eqnarray}
  They differ from the previous equations (\ref{cir.23xx}) only for the sign of the time evolution.
  It follows that the set of periodic solutions, described in the previous analysis for the attractive traps, still exist in the present case of repulsive traps, \emph{with the same periods}. Only the shape of the trajectories is affected by replacing diverging exponential in place of converging ones.\\

\begin{figure}[t!]
\centering
\includegraphics[bb = 150 5 180 200, height=42mm]{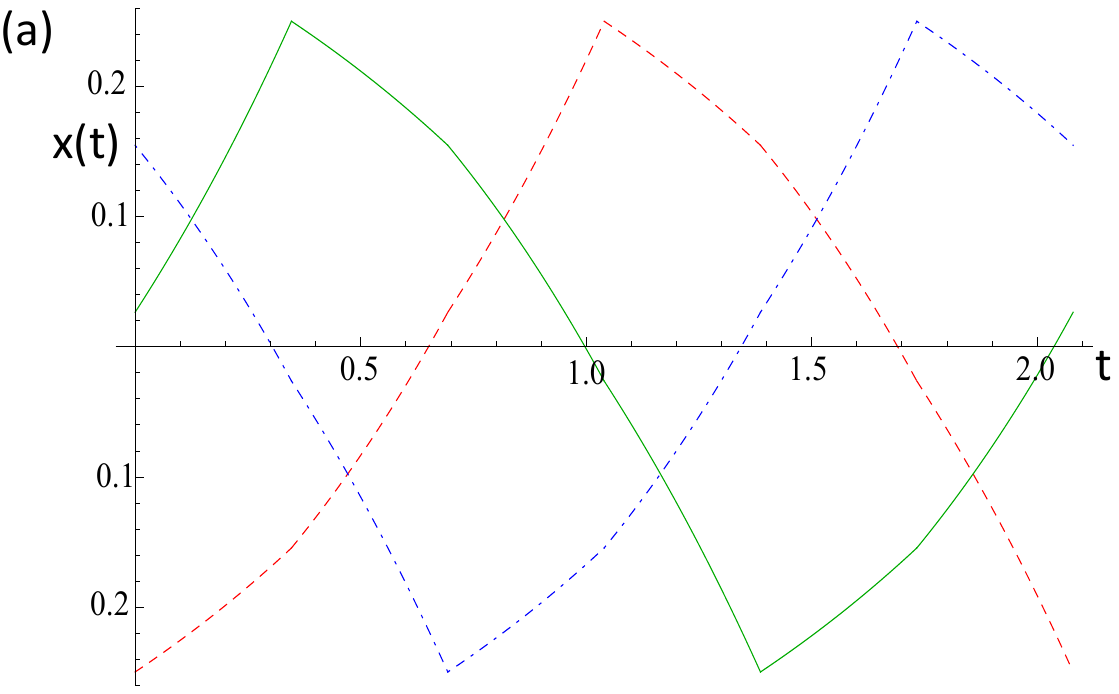}\\
\includegraphics[bb = 130 5 180 200, height=40mm]{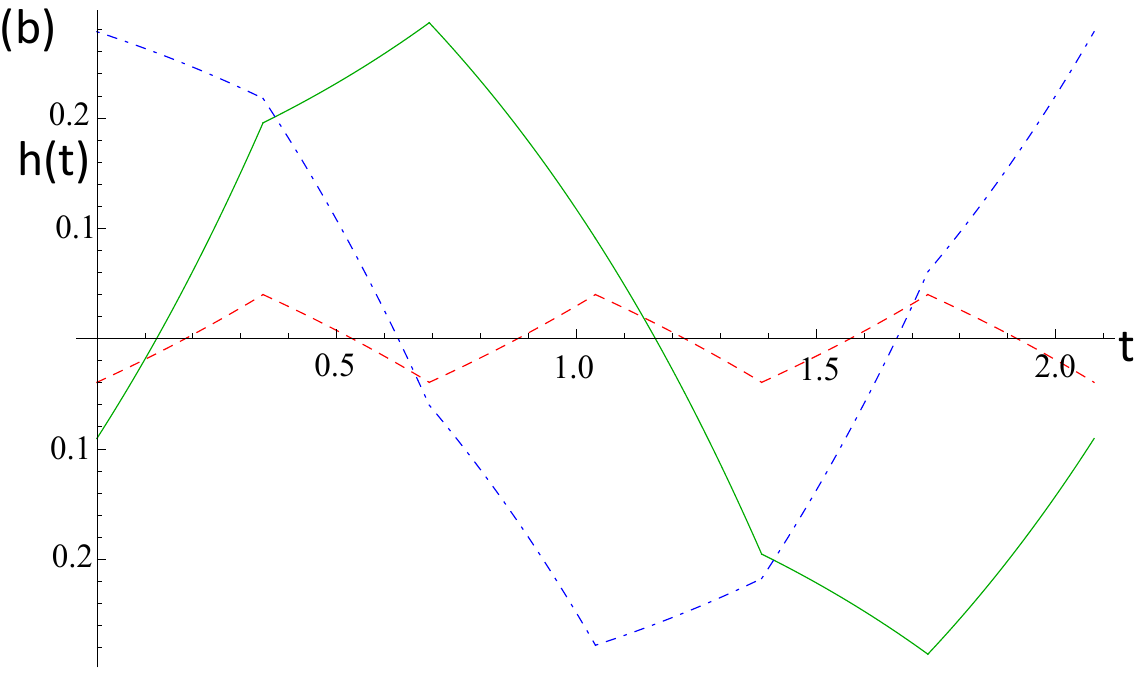}\\
 \caption{(Color online) (a) The trajectories for the the phase-locked solution for three particles in the {\it repulsive} case are plotted versus time, in units $\tau_0$. (b) The normal modes $h_1(t)$ in red-dash, $h_2(t)$ in green-solid,  $h_3(t)$ in blue dash/dot are plotted versus time, in units of $\tau_0$.}
  \label{fig12}
\end{figure}

Fig.~\ref{fig12}a and Fig.~\ref{fig12}b  are examples of this point, when compared with the Fig.~\ref{fig2}a and Fig.~\ref{fig2}b. We plotted the phase-locked solution for three particles in the repulsive case. Fig.~\ref{fig12}a shows the trajectories of the three particles  as function of time, for one period $T=6\Delta$.
 Fig.~\ref{fig12}b plots the normal modes $h_1(t)$, $h_2(t)$  and $h_3(t)$, again for one period. The oscillations of the pair of normal modes corresponding to the degenerate eigenvalue have an amplitude much bigger then the oscillations of the normal mode $h_1(t)$.  Remarkably, the simulations  show that in the case of repulsive traps it is this periodic solution, the one with the \emph{shortest} period, which is the stable asymptotic synchronization for the system.\\
 
With systems of higher $n$, we have observed in the numerical simulations that again on switching from attractive to repulsive traps the stable solution ceases to be that with longest period. Instead, the system converges onto one of the solutions with shortest period. However, as is clear by observing the sequence of solution periods $T$ calculated for the cases ($n=4$ or $n=5$),  while there is a clearly separate largest $T$, there are various solutions with small $T$.  We have tested numerical solutions in the physically realistic parameter space, using the two numerical approaches outlined in section~\ref{simulations}: the ODE solver with the ``strong'' tangential motion constraint converges to the shortest period solution, whereas the Brownian Dynamics simulation with the weaker ``harmonic'' constraint converges to one of the shorter period solutions, but not always the shortest one.  This will be explored further in future work; our current understanding is that the system with repulsive potentials converges to one of the solutions with shorter periods, and that when there are various periods grouped close together the details of the system play a strong role in selecting between these.

    \section{Conclusion}

      The model studied in this paper has several unusual features.
    Most studies on synchronization show that it is a threshold process: it occurs when the coupling among the oscillators is strong enough or when increasing the number of oscillators, a large fraction synchronize to a common frequency~\cite{pikovsky01}. In contrast, in the present model, where we studied the deterministic system without stochastic forces, we find no threshold.  Synchronization occurs for any strength of the hydrodynamic coupling, that is any value of $a/R$, and any number of oscillators.\\
     Most models on synchronization are described by  non-linear differential equations where a very limited progress is achieved analytically, whereas the present model is piece-wise linear and all periodic solutions may be derived.\\
     Numerical solutions in the case of {\it attractive} harmonic potentials clearly show the synchronization to the periodic solution with the longest period, for any number of particles.       In the case of {\it repulsive} harmonic potentials, numerical simulations show the synchronization to the
     the pair of phase locked solutions in the case of three particles.  We have not performed a stability analysis about the periodic solutions, but there is no reason to expect disagreement with  the results of numerical simulations.
     This change of the asymptotic solution with the change of curvature of the potential had already been shown numerically for a linear chain in  \cite{stark11}.
     The analytical framework developed here  adequately describes  experiments on systems of colloidal particles with  moving optical traps~\cite{cicuta11z}. \\

 \section{Appendix}
 In the case of three particles, with attractive harmonic traps, the initial conditions for the normal modes, for periodic phase-locked solutions are given in the eqs.(\ref{e.2}). We  supplement them with  an initial condition:
 $$x_1(t_0)=\frac{1}{\sqrt{3}}h_1(t_0)-\frac{2}{\sqrt{6}}h_3(t_0)=-\frac{\lambda}{2}+\xi,$$
 and obtain an equation that determines $\Delta$:
 \begin{eqnarray}
 &&\frac{\lambda}{6}\left(1-e^{-\Delta/\tau_1}\right) \left(  e^{-2\Delta/\tau_1}-e^{-\Delta/\tau_1}+1\right)\left(1+e^{-3\Delta/\tau_2}\right)
\nonumber\\ &&
+\frac{\lambda}{3}\left(1-e^{-\Delta/\tau_2}\right)  \left(  e^{-2\Delta/\tau_2}+2e^{-\Delta/\tau_2}+1\right) \left(1+e^{-3\Delta/\tau_1}\right)\nonumber\\
&&=\left(\frac{\lambda}{2}-\xi\right)\left(1+e^{-3\Delta/\tau_1}\right)\left(1+e^{-3\Delta/\tau_2}\right). 
  \label{e.88}
 \end{eqnarray}
 Furthermore, ${\tau_0}/{\tau_1}+2{\tau_0}/{\tau_2}=3$,
 since $\lambda_1+2\lambda_2=0$.
 Let us define:
  $$x=e^{-\Delta/\tau_0} , \,\,   p_1=\frac{\tau_0}{\tau_1}<1 ,\,\,   p_2=\frac{\tau_0}{\tau_2}>1
 ,\,\,   p_1+2p_2=3, $$
 then Eq.(\ref{e.88})  may be rewritten in the simpler form:
  \begin{eqnarray}
 &&\frac{\lambda}{3}\Big[ -x^{p_1}+x^{2p_1} -x^{3p_1}+x^{p_2}\left(1+x^{p_1}-x^{2p_1}+2x^{3p_1}\right)+\nonumber\\
&&\,\,\,\, +\,x^{2p_2}\left(-2-x^{p_1}+x^{2p_1}-3x^{3p_1}\right)\Big]+
 \nonumber\\
 &&\,\,\,\,+\,\xi\left(1+x^{3p_1}\right)\left(1-x^{p_2}+x^{2p_2}\right)\,=\,0.
    \label{e.19}
   \end{eqnarray}
   If we choose the coupling strength ${3a}/{(8R)}=\sqrt{3}/5$, we have $p_1=0.8$ and $p_2=1.1$. The further behaviour of $\Delta$ as function of $\lambda/\xi$ is depicted in Fig.~\ref{fig4}.\\
   If we further choose
     $\xi=\lambda/4$, then eq.(\ref{e.88}) implies
      $x \sim 0.70701565$, that is:
   $$\frac{\Delta}{\tau_0}=-\log x \sim 0.346702 , \,\,\,
   T=6\Delta=-6\tau_0 \log x \,.$$

  For the case of five particles with attractive traps,
  Fig.~\ref{fig14} shows the period $T$ in units of $\tau_0$ for the phase-shifted solution  with adjacent trajectories $\{1,3,5,2,4 \}$,
 the period  of the phase-shifted solution  with adjacent trajectories $\{1,2,3,4,5 \}$, and the in-phase solution, versus the hydrodynamic coupling $3a/(8R)$, while we fixed $\lambda=1$, $\xi=0.25$.
  In the limit of vanishing coupling, the three curves converge to $T=2\tau_0 \log (\lambda-\xi)/\xi \sim 2.1972\,\tau_0 $.\\

{\bf Acknowledgments}\\
We are grateful for many discussions with Filippo de Lillo, Nicolas Bruot and Romain Lhermerout.

\bibliography{BIBDATAv41}

\end{document}